\documentstyle[aps,floats,epsf,twocolumn]{revtex}
 
\begin{document} \draft

\title{Quantum critical points with the Coulomb interaction and the
dynamical exponent: when and why $z=1$ }

\author{Igor F. Herbut}

\address{Department of Physics, Simon Fraser University, 
Burnaby, British Columbia, \\
 Canada V5A 1S6\\}\maketitle

\begin{abstract}
A general scenario that leads to Coulomb quantum criticality with  the 
dynamical critical exponent $z=1$ is proposed. I point out that the 
long-range Coulomb interaction and quenched disorder have competing
effects on $z$, and that balance between the two may lead to charged
quantum critical points at which $z=1$ exactly. This is illustrated with
the calculation for the Josephson junction array Hamiltonian in dimensions
$D=3-\epsilon $. Precisely in $D=3$, however, the above simple result
breaks down, and $z>1$. Relation to other studies is discussed. 
\end{abstract}

\vspace{10pt}

The crucial difference  between the quantum ($T=0$)
and the more familiar finite temperature phase transitions is that
while dynamics is irrelevant for the latter, it is essential for the
former \cite{hertz}. The link between statics and dynamics at a
continuous quantum phase transition is usually
parameterized with the value of {\it dynamical} critical exponent $z$ 
that describes relative scaling of the time and the length scales in the
problem \cite{sondhi}, \cite{subir}.
Together with the correlation length exponent $\nu$,
$z$ enters the low-temperature scaling of all physical
observables, since in the vicinity of a quantum critical point
temperature scales as
$T\sim |\delta| ^{z\nu}$, where $\delta$ is the $T=0$ tuning parameter.
The value of $z$ for a given quantum critical point is therefore
of great interest, and has often been used to distinguish one
universality class from another.
In their seminal paper on universal conductivity in two dimensions, 
Fisher, Grinstein and Girvin \cite{girvin} also proposed that
when the long-range Coulomb interaction is present, at the criticality
energy should scale as inverse of length, so that $z=1$ should 
result. This well-known conjecture has since been used in interpreting
some of the most intriguing experiments in modern condensed matter
physics, ranging from superconductor-insulator
transitions in low-dimensional systems \cite{goldman}, via 
metal-insulator  transitions in Si-MOSFET's \cite{abrahams}, to
the universality of the underdoped high-$T_c$ cuprates \cite{schneider}.
It has also been utilized in the Monte Carlo simulations where knowing
$z$ in advance greatly simplifies the inevitable finite-size scaling
analysis \cite{wallin}.

The purpose of this Letter is to provide the theoretical
justification for this widely used relation and point to its limitations.
I show that in certain dimensions Coulomb
coupling constant (i. e. charge) is protected from
renormalization, and consequently its flow under scaling transformation
is directly proportional to its canonical dimension,  which is just 
$z-1$ \cite{girvin} (see Eq. (2)). This
implies that {\it if} a charged critical points exists in the 
theory, its location is actually determined by the solution of the
{\it equation} $z=1$, which then also determines the value of dynamical
critical exponent exactly. I argue that this
situation arises when there is an additional coupling in the theory with
the competing effect on the dynamical exponent, and which
can balance the effect of Coulomb interaction.
Such a coupling is shown to be provided by quenched disorder,
and a concrete realization of the above scenario is
worked out on the example
of a disordered Josephson junction array Hamiltonian in $D=3-\epsilon$
dimensions. Finally, the simple relation 
$z=1$ is found to break down at special dimensions at which the
above renormalization group (RG) protectorate on charge is lifted.
Relation to other recent theoretical studies of Coulomb criticality
is discussed.

To be specific, I will focus on the theory originally considered in
\cite{grinstein} which describes an array of coupled Josephson junctions,
but it will transpire that the underlying physics is 
more general. Building on an earlier work by Ma \cite{ma}, 
Fisher and Grinstein \cite{grinstein} have shown that Coulomb
interaction can be represented by a minimal coupling to the soft
{\it scalar} gauge field. In the long wavelength limit the critical
field theory that describes the system of bosons interacting via Coulomb
interaction at $T=0$ and in $D$-dimensions takes the form \cite{grinstein}
\begin{eqnarray}
S=\int d^D \vec{r} d\tau [  |(\partial_\tau- i A_0) \Psi|^2 +
|\nabla \Psi|^2 + \nonumber \\
(V(\vec{r},\tau)+ m^2) |\Psi|^2
+ \lambda |\Psi|^4 + \frac{1}{2e^2} A_0 |\nabla|^{D-1} A_0 ], 
\end{eqnarray}
where by $|\nabla|^{D-1}$ it is meant $|\vec{k}|^{D-1}$ in the Fourier space.
$\Psi(\vec{r},\tau) $ is the complex superfluid order parameter,
and $A_0(\vec{r},\tau)$ is the scalar gauge field, which when integrated
out introduces the Coulomb interaction into the remaining action
for $\Psi$.
I have also included a random potential $V(\vec{r},\tau)$, more on which
in a moment. More generally, the gauge-field propagator in the momentum
space is
$(e^2 (V_c (k) -1))$ \cite{grinstein}, which for the Coulomb interaction
$V_c (\vec{r}) = 1/r$ then yields the last term in (1) at small momenta. The
presence of a neutralizing background is included by omitting  the 
$k=0$ components of the gauge field \cite{ma}.

The above action without randomness ($V(\vec{x},\tau)\equiv 0$) was
first studied by Fisher and Grinstein \cite{grinstein},
and recently revisited by Ye \cite{ye}. In $D=3$ both the charge ($e^2$)
and the quartic interaction ($\lambda$)  are marginally
irrelevant, and the Coulomb interaction causes the flow to run away to
negative $\lambda$, which may be interpreted as a sign of a
discontinuous transition. In $D=3-\epsilon $ the result is more interesting:
charge is irrelevant at the $XY$ critical point, but if too large may still
lead to a runaway flow. Irrelevance of the charge in the theory
(1) arises in a somewhat non-trivial way, and it will prove instructive
to understand this in some detail.
In $D<3$ the inverse gauge-field propagator in (1) is
{\it non-analytic} in $k$ and
therefore $e^2$ can not get renormalized by the integration over the
high-energy modes. This RG protectorate is reminiscent of the situation in the
$2+1$-dimensional electrodynamics with the Chern Simons term
\cite{semenoff}, where the statistical angle is exactly marginal
for a similar reason. Here, this  
means that the $\beta$-function for the charge
is determined exclusively by its canonical dimension. Gauge invariance and
the accompanying Ward identity imply that $A_0 \sim \tau^{-1}$, so
assuming $\tau\sim L^z$ yields to  
\begin{equation}
\frac{ d e^2}{d \ln(b)}  = e^2 (z-1)
\end{equation}
in $D<3$, where $b$ is the standard RG parameter.
The exponent $z$ in the last equation then needs to be determined
from the renormalization of the $\Psi$-propagator. The simplest
one-loop calculation (see the first diagram on Fig. 1 a) )
in $D=3-\epsilon$ then gives
$z= 1- e^2/3$, and thus smaller than one. By the Eq. (2) small charge is
then irrelevant.
Two features of this result that are likely to be quite 
general should be noticed: first,
$z\neq 1$ since the scalar gauge-field
couples only to the time derivative, and thus discriminates
between space and time. Secondly, the {\it negative} sign in
the Eq. (2) comes from the tendency of the gauge-field to {\it soften} the
$\Psi$-propagator. It is analogous to the well-known (and much debated) 
result that the anomalous dimension in the scalar (Higgs) electrodynamics
is negative \cite{herbut}, \cite{sudbo}, \cite{kleinert}. It could
therefore be expected that when higher order terms in $e^2$  are
included $z<1$ will remain, and that irrelevancy of the charge
in the pure theory (1) is more general than the simple one-loop calculation
would suggest.

The Eq. (2) also implies that if there would exist a critical point in
the theory with non-zero charge, in $D<3$ at that fixed point $z=1$ exactly. This
non-perturbative result parallels another  exact result $\eta_A=1$ in the
$D=3$ scalar electrodynamics \cite{herbut}, \cite{teitel}
where $\eta_A$ is the anomalous dimension of the vector gauge-field.
The crucial question is
how can such a Coulomb critical point (with $e^2 \neq 0$) arise.
A clue is provided already in the above discussion: one needs
another coupling  which will tend to increase $z$ and balance the
effect of Coulomb interaction in the Eq. (2).
A physically realistic candidate is disorder: in the 
quantum theory the random potential
$V(\vec{r},\tau)$ is random in space but static and
independent of imaginary time. This anisotropy will in general lead to
a non-trivial $z$,  and it is well-known that the effect to the lowest order
is always to {\it increase} it \cite{herbut1}. In other words, while the
weak Coulomb interaction is marginally irrelevant at the XY critical point, I  
expect it to become {\it relevant} at the disordered critical point 
where $z>1$. Small charge near the random critical point
should grow until it balances disorder in the Eq. (2). As a
result, a new stable (Coulomb) critical point may arise, at
which $z=1$ exactly.

Next I demonstrate that the above scenario is indeed born out 
in the one-loop calculation in the theory (1), in passing reconciling
the apparently different results in refs. \cite{grinstein} and \cite{ye}.
I then proceed to show how the equality $z=1$ is violated in $D=3$,
and comment on relations to other works. 

\begin{figure}[t]
\epsfysize 11cm
\epsfbox[0 -40 250 700]{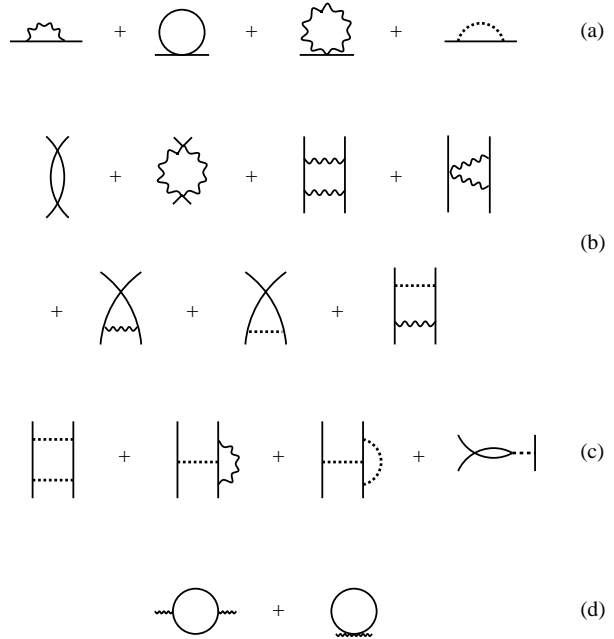}

\noindent
\caption[]{ One loop contributions to the renormalization of self-energy,
quartic interaction, disorder coupling, and polarization. Wavy lines
represent the gauge-field propagator, and dashed lines the disorder
vertex.}
\label{f1}
\end{figure}

To exert some control over the fixed points in the theory I will assume both a
small $\epsilon =3-D$, and $\epsilon_\tau$, where 
the latter is the number of dimensions over which
disorder is correlated \cite{cardy}. The physically interesting case 
correspond to $\epsilon_\tau=1$, but since I am primarily interested
in the point of principle, convergence  properties 
of the double-$\epsilon$ expansion will not be of much concern here 
\cite{mukho}. To average over
disorder I utilize the standard replica trick, which attaches 
a replica index $\alpha = 1, ...N$ onto all fields in the action
(1) and introduces another interaction-like
term in the theory: 
\begin{eqnarray}
-\frac{W}{2} \sum_{\alpha, \beta=1}^N \int d^{D+1}\vec{R} d^{D+1} \vec{R}'
\delta^{D+1- \epsilon_\tau} ( \vec{R}-\vec{R}') \nonumber \\
|\Psi_\alpha (\vec{R})|^2 |\Psi_\beta (\vec{R}')|^2, 
\end{eqnarray}
where the limit $N\rightarrow 0$ is to be taken
at the end of the calculation \cite{cardy}. Here
$\vec{R}= (\vec{r},\tau)$, and for $\epsilon_\tau =1$ one recovers the
quantum problem where disorder is uncorrelated (Gaussian) in space
and independent of the imaginary time.

  To perform the Wilson-Fisher momentum shell renormalization group
one integrates out all the fields with $\Lambda /b <k< \Lambda$
and $-\infty<\omega<\infty$, where
$\Lambda$ is the ultraviolet cutoff. The effect
of this procedure is to alter  the 
coefficients in front of $\omega^2$ ($Z_\omega$), $k^2$ ($Z_k$),
$m^2$ ($Z_m$), $\lambda$ ($Z_\lambda$),
$W$ ($Z_W$), and $A_0^2$-term ($Z_A$)
in the Fourier transformed action (1) \cite{remark}.
One then rescales the momenta and the frequencies as $b k\rightarrow k$ and
$b^z \omega \rightarrow \omega$, and the fields as
$b^{-D} A_0 \rightarrow A_0$ and $b^{-(2+D+z)/2}Z_k^{1/2} \Psi
\rightarrow \Psi$, to find finally that by defining new coupling constants
as $\lambda(b) = b^{4-D-z} Z_k ^{-2} Z_\lambda \lambda$,
$ W(b) = b^{4-D-z+\epsilon_\tau} Z_k^{-2} Z_W  W$,  and
$e^2 (b) = b^{z-1} Z_A ^{-1} e^2$ the action can be restored into
its original form (at the critical point $m^2 =0$).
Computing next the $Z$-factors
diagrammatically to one-loop order (Fig. 1) gives at the criticality
\begin{equation}
\frac{de^2}{d \ln(b)} = e^2 (z-1) -\delta_{3,D} \frac{1}{12}e^4, 
\end{equation}
 \begin{equation}
 \frac{d\lambda}{d \ln(b)} = (\epsilon + \frac{1}{2}e^2 + \frac{11}{8}W )\lambda
 -\frac{5}{2}\lambda^2  - \frac{1}{4}e^4 , 
 \end{equation}
 \begin{equation}
 \frac{d W}{d \ln(b)}
  = (\epsilon+\epsilon_\tau - 2\lambda +\frac{1}{2}e^2) W +\frac{7}{8}W^2 . 
 \end{equation}
 The exponent $z$  in the Eq. (4)
 is determined by demanding that $b^{-2z} Z_\omega = b^{-2} Z_k$, which gives
 \begin{equation}
 z= 1+ \frac{1}{8}W - \frac{1}{3} e^2 .
 \end{equation}

Note that for $D<3$ the flow equation for the charge reduces to the Eq. (2).
Precisely in $D=3$ the inverse
propagator for the gauge field becomes analytic,  $\sim k^2$. This
means that in $D=3$
the charge becomes renormalized by the polarization diagrams
in Fig. 1 d), which contribute the last term in the Eq. (4).
For $W=0$ the above $\beta$-functions reduce to those of ref. 
\cite{grinstein} when $D<3$, and coincide with those of \cite{ye} right
at $D=3$, upon simple redefinitions of couplings. They are also
equivalent to those of \cite{mukho} for $e^2=0$. It may be interesting to note
that many of the individual diagrams on Fig. 1 are ultraviolet divergent,
due to the independence of the gauge-field propagator on frequency. 
All those divergences exactly cancel out in the final result
\cite{ye}. Finally, the flow of the mass term in (1) yields the correlation
length exponent:
\begin{equation}
\nu = \frac{1}{2}+ \frac{1}{4}(\lambda+ \frac{e^2}{6} - \frac{W}{4}).
\end{equation}

   Let us turn now to the fixed points of the above equations. Besides
the Gaussian and the XY fixed points at $W=e^2 = 0$, both unstable with
respect to disorder, there are two disordered fixed points. 
First, at $e^2 =0$, there is a {\it neutral} 
disordered fixed point \cite{cardy} at
$\lambda_n = 2(4\epsilon +11 \epsilon_\tau)/9$ and
$W_n = 8(\epsilon+ 5\epsilon_\tau)/9$, which is attractive in the $\lambda-W$
plane. At the neutral fixed point $z_n =1+W/8 >1$, so a weak Coulomb
interaction is a relevant perturbation. With a small charge turned on,
the flow is towards a new, stable, Coulomb critical point. In $D<3$ this
fixed point is located at
\begin{eqnarray}
\lambda_c = \frac{100}{183} \epsilon \{ 1+ \frac{389 \epsilon_\tau}
{100\epsilon} + \nonumber \\
\sqrt{ (1+ \frac{389\epsilon_\tau}{100\epsilon})^2 +
\frac{1647}{5000} (1+\frac{\epsilon_\tau}{\epsilon})^2 }   \}, 
\end{eqnarray}
\begin{equation}
W_c = \frac{16}{17} ( 2 \lambda_c -\epsilon-\epsilon_\tau), 
\end{equation}
and
\begin{equation}
e^2 _c = \frac{3}{8} W_c, 
\end{equation}
where the last equation ensures that $z=1$. The reader should note that
disorder is necessary for the existence of the Coulomb critical point:
without it the critical point would turn imaginary, and
one would find only the standard runaway flow 
characteristic of the gauge-field fluctuations in the $\epsilon$-expansion
\cite{halperin}. The Coulomb critical point therefore may
be considered as an example of a disorder {\it induced}
continuous phase transition.

 Precisely in $D=3$ the last term in the
Eq. (4) becomes finite. There still exists a stable Coulomb critical point
at $\lambda_c = 3.61 \epsilon_\tau$, $W_c = 40( 2\lambda_c -\epsilon_\tau)/41$,
and $e^2 _c = 3 W_c /10$, but with the dynamical critical exponent 
\begin{equation}
z= 1+ \frac{e^2 _c}{12}, 
\end{equation}
which gives $z\approx 1.15$, for example, for
$\epsilon_\tau =1$. One finds $z\neq 1$ in $D=3$ as a result
of the removal of the RG protectorate on charge. The same violation of
the simple relation $z=1$ can be expected in other problems in special
dimensions. Also, since at the criticality $V_c (r) \sim 1/r^z$, one expects
that $z\geq 1$ in general \cite{girvin}, since screening should certainly
not make the
interaction longer ranged. The result (12) thus implies the Coulomb 
interaction has been partially screened at the criticality in $D=3$, and now
decays faster (but still as a power-law) with distance \cite{note}. 

The result $z=1$ was previously also found in the large-N theory of Dirac
fermions in $D=2$ \cite{sachdev}, interacting both via Coulomb interaction
and with the Chern Simons field. This theory
may be relevant to the quantum Hall
state-insulator transition \cite{sondhi}, \cite{kivelson}.
It was found that for a certain range
of the statistical angle (Chern Simons coupling) the competition
between the Chern Simons and the Coulomb interaction leads to a non-trivial
charged fixed point, at which $z=1$ exactly, just like in the
Eq. (2). The physical reason why the two interactions
have competing effects on $z$ only for some values of the Chern Simons
coupling remained somewhat obscure in this work.
Nevertheless, the result in \cite{sachdev} bears a formal resemblance to mine.

  Finally, $z=1$ was also found in the previous work by the author
on the quantum critical behavior of dirty bosons with Coulomb
interaction in $D=1+\epsilon$ dimensions \cite{igor}. There it arises
as a consequence of a special symmetry the theory dual to (1)
possesses precisely in $D=1$, and can therefore be suspected to be
an artifact of the specific RG scheme that was employed. Present paper 
can thus be understood as complementing the previous work
in that it shows that $z=1$
is also exact near the $D=3$, and it may therefore be expected 
to hold in the physical case $D=2$.

It may also be interesting to note that simply setting $\epsilon=\epsilon_\tau
=1$ in the one-loop Eqs. (7) and (8), to crudely estimate the exponents
in $D=2$, besides the exact $z=1$ also yields $\nu= 1.46$ at the Coulomb
criticality. Experimentally, $\nu\approx 1$ \cite{goldman}, and the
result for $\nu$ is less reliable than the one obtained in the expansion
around $D=1$ \cite{igor}.

To summarize, it was shown that the competition between
the correlated (quantum) disorder and Coulomb interaction may lead to
new charged critical point at which the dynamical critical exponent
$z=1$ exactly. This simple result breaks down in special dimensions,
and an example of the Josephson junction array when this 
happens in $D=3$ was provided. I argued that similar results 
should be expected whenever there are two couplings in
the theory with competing effects on $z$.

The author is grateful to Matthew Fisher and Shivaji Sondhi for
their useful comments and interest in this work.
This research was supported by NSERC of Canada and the Research Corporation.


\begin{thebibliography}{99}
\bibitem{note} Gauge invariance prohibits a mass term for the gauge-field
in Eq. (1) that would yield the  Yukawa form of the screened
interaction in real space.
\end{thebibliography}
\end{document}